\documentclass[10pt,letterpaper]{article}
\usepackage{opex3}
\usepackage{amsmath,amssymb}        
\usepackage[mathscr]{euscript}      
\usepackage{dsfont}                 
\usepackage{graphicx}

\usepackage{color} 

\begin{document}

\title{Diffractively coupled Fabry-Perot resonator with power-recycling}

\author{Michael Britzger$^1$, Daniel Friedrich$^1$, Stefanie Kroker$^2$, Frank Br\"uckner$^2$, Oliver Burmeister$^1$, Ernst-Bernhard Kley$^2$, Andreas T\"unnermann$^2$, Karsten Danzmann$^1$, and Roman Schnabel$^{1,*}$}
\address{$^1$Albert-Einstein-Institut, Max-Planck-Institut f\"ur Gravitationsphysik and Leibniz Universit\"at Hannover, Callinstr. 38, 30167 Hannover, Germany\\$^2$Institut f\"ur Angewandte Physik, Friedrich-Schiller-Universit\"at Jena, Max-Wien-Platz 1, 07743 Jena, Germany\\}
\email{$^*$roman.schnabel@aei.mpg.de}


\begin{abstract}
We demonstrate the optical coupling of two cavities without light transmission through a substrate. Compared to a conventional coupling component, that is a partially transmissive mirror, an all-reflective coupler avoids light absorption in the substrate and therefore associated thermal problems, and even allows the use of opaque materials with possibly favourable mechanical and thermal properties. Recently, the all-reflective coupling of two cavities with a low-efficiency 3-port diffraction grating was theoretically investigated. Such a grating has an additional (a third) port. However, it was shown that the additional port does not necessarily decrease the bandwidth of the coupled cavities. Such an all-reflective scheme for cavity coupling is of interest in the field of gravitational wave detection. In such detectors light that is resonantly enhanced inside the so-called power-recycling cavity is coupled to (kilometre-scale) Fabry-Perot resonators representing the arms of a Michelson interferometer. In order to achieve a high sensitivity over a broad spectrum, the Fabry-Perot resonators need to have a high bandwidth for a given (high) power build-up. We realized such an all-reflective coupling in a table-top experiment. Our findings are in full agreement with the theoretical model incorporating the characteristics of the 3-port grating used, and therefore encourage the application of all-reflective cavity couplers in future gravitational wave detectors.
\end{abstract}



\section{Introduction}
Over the past several years an international network of laser-interferometric gravitational wave (GW) detectors has been built 
[1--3]. The detectors have reached sensitivities at or close to their design sensitivities; however, the first detection of a gravitational wave has not yet been achieved. Sensitivity upgrades of existing detectors are in progress 
[4--6] while new detectors such as the Japanese LCGT are being built \cite{LCGT2010}, a southern hemisphere detector AIGO has been proposed \cite{AIGO2010} and investigations towards the third generation like the European Einstein Telescope are being carried out \cite{ET2010}. 

\begin{figure}[htpb]
\centerline{\includegraphics[height=4.03cm]{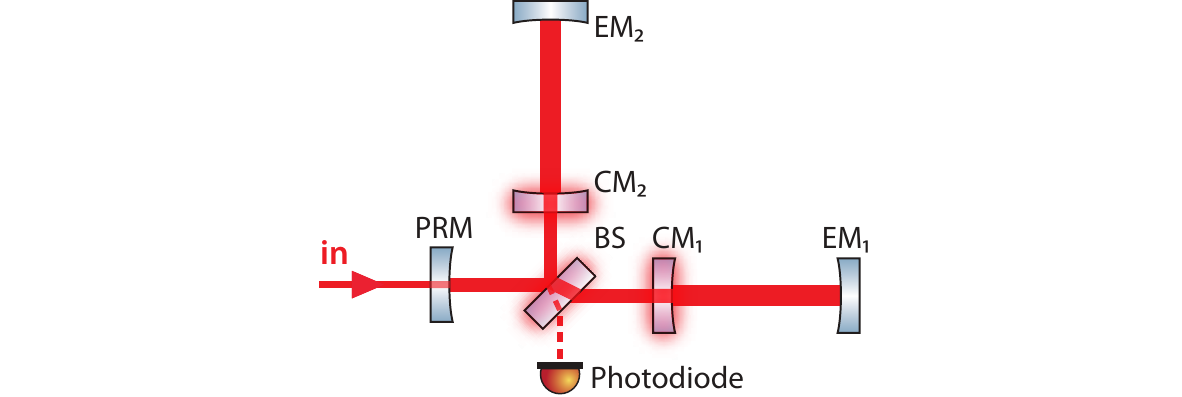}}\label{Fig:IFO_heat}
 \caption{Advanced Michelson-type interferometer with power-recycling cavity and arm resonators. The transmissive beamsplitter (BS) and the coupling mirrors to the arm cavities (CM$_\mathrm{n})$ are exposed to high thermal load (red blur). Each arm cavity (coupling mirror CM$_\mathrm{n}$ and end mirror EM$_\mathrm{n}$) together with the power-recycling mirror (PRM) forms a system of two coupled cavities.}
\end{figure}

Present and also planned GW detectors are based on Michelson interferometers, which incorporate coupled optical cavities to improve their shot noise limited sensitivity. Fabry-Perot arm resonators coupled to a so-called power-recycling resonator provide very high light power build-ups inside the interferometer without decreasing the bandwidth too much. Figure~\ref{Fig:IFO_heat} shows such an interferometer layout 
[10--14]. For future detectors with further increased light power build-ups the optical absorption of the transmissive optics might become a sensitivity limiting problem. Light absorption in the substrates causes thermal effects like thermal lensing and photo-thermo-refractive noise 
[15--17], and will therefore lead to an upper limit for the circulating power. This concerns the balanced beamsplitter and the coupling mirrors to the interferometer arm resonators (see Fig.~\ref{Fig:IFO_heat}). To reduce the absorption in substrates all-reflective technologies have been proposed and realized 
[18--21]. The proposals rely on replacing the transmissive components with nano-structured dielectric reflection gratings that avoid any light transmission through the substrate. As another profiting aspect, all-reflective technologies allow the use of opaque materials with favourable mechanical and thermal properties \cite{Nawrodt07}. In \cite{gratingbs} the all-reflective replacement of the balanced beamsplitter of a Michelson-type interferometer with an appropriate diffraction grating \cite{fahr2007} was demonstrated experimentally. All-reflective cavity couplers proposed can be divided into 2-port-couplers operated in first-order Littrow arrangement, as realized in \cite{Sun,Bunki06}, and 3-port-couplers operated in second-order Littrow arrangement, as realized in \cite{bunkowski2}. The latter concept introduces an additional port that potentially acts as an additional loss channel; however, it relies on only low diffraction efficiencies and correspondingly shallow grating structures, which suggests that cavities with well-defined Gaussian TEM00 modes and very high finesse values are feasible.

Here, we experimentally demonstrate the all-reflective coupling of a Fabry-Perot resonator with a power-recycling resonator. The coupling component is a dielectric low-efficiency 3-port diffraction grating. This scheme was theoretically analyzed in \cite{burmeister2010}. The 3-port grating was characterized with respect to diffraction efficiencies and losses. Based on the characterization and the analysis in \cite{burmeister2010} we simulated the light powers at all three output ports of the coupled cavity system in dependence of the coupled cavity detunings. We find very good agreement between the experimental observation and our model.

In the following we first start with a brief theoretical description of the coupled cavity system and also discuss our interferometer topology and requirements for the mode matching of the elliptical beam profiles involved.

\section{3-port grating coupled cavities}

\begin{figure}[htpb]
\centerline{\includegraphics[height=3.05cm]{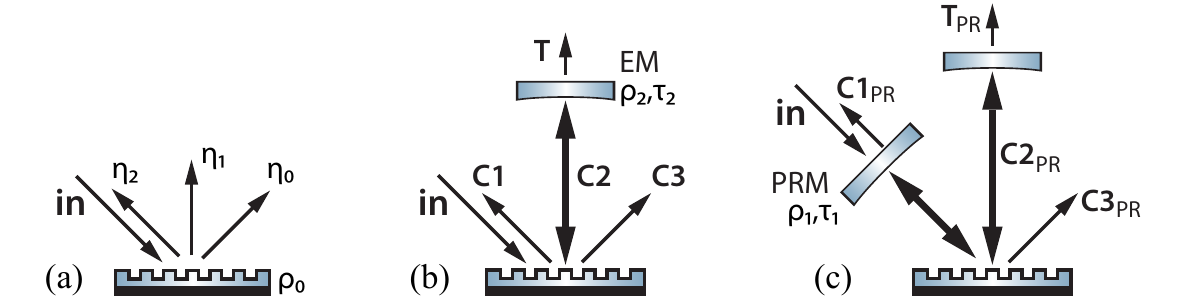}}\label{Fig:3PGtoPR}
 \caption{(a) 3-port grating in second-order Littrow mount. The amplitude diffraction efficiencies are denoted $\eta_n$ and reflectivity under normal incidence $\rho_0$, respectively. (b) By inserting a mirror (EM) perpendicular to the grating a 3-port-grating cavity is generated, with the diffraction efficiency $\eta_1$ as the coupling efficiency. The light fields are labeled as follows: The backwards reflected field towards the laser source C1, the intra-cavity field C2, the forward reflected field at the additional grating port C3 and the transmitted field T. (c) Inserting the power-recycling mirror PRM in the entrance, a power-recycled 3-port grating coupled cavity is formed. The light fields are denoted C1$_\mathrm{PR}$, C2$_\mathrm{PR}$, C3$_\mathrm{PR}$ and T$_\mathrm{PR}$.}
\end{figure}

In contrast to a conventional coupling mirror, the 3-port grating couples an incoming beam into three instead of two output ports. Figure \ref{Fig:3PGtoPR}~(a) shows a 3-port grating mounted in second-order Littrow configuration. The amplitude diffraction efficiencies of each diffraction order are denoted $\eta_n$ and the amplitude reflectivity under normal incidence $\rho_0$, respectively. The arm cavity is then established perpendicular to the grating surface with a small first order diffraction efficiency as the coupling efficiency to the cavity [see Fig.~\ref{Fig:3PGtoPR}~(b)].

The back- and forward-reflected ports are denoted C1 and C3. In \cite{bunkowski1} the phase relations for a binary structured grating without asymmetry (blaze) were derived for the lossless case. For a Fabry-Perot cavity with a 3-port grating as the coupling component as shown in Fig.~\ref{Fig:3PGtoPR}~(a) the amplitude reflection coefficients for the back-reflected field $c_1$, the forward-reflected field $c_3$ as well as the amplitudes for the intra-cavity field $c_2$ and the transmitted field $t$ are described as
\begin{eqnarray}
c_1&=&\eta_2\exp{(\mathrm{i}\phi_2)}+\eta_1^2\rho_2\exp{\left[\mathrm{i}2(\phi_1+\Phi_2)\right]}d_2,\label{dreiport/c1}\\
c_2&=&\eta_1\exp{(\mathrm{i}\phi_1)}d_2,\label{dreiport/c2}\\
c_3&=&\eta_0+\eta_1^2\rho_2\exp{\left[\mathrm{i}2(\phi_1+\Phi_2)\right]}d_2,\label{dreiport/c3}\\
t&=&\mathrm{i}\tau_2\exp{(\mathrm{i}\Phi_2)}\,c_2,\label{dreiport/t}
\end{eqnarray}
where the amplitude reflectance and transmittance of the cavity end mirror as the second component are given by $\rho_2$ and $\tau_2$, respectively. The resonance factor is given by $d_2=[1-\rho_0\rho_2\exp{(\mathrm{i}2\Phi_2})]^{-1}$, and the length $L_2$ of the grating cavity is expressed by the detuning parameter $\Phi_2=\omega{L_2/c}$, where $\omega$ is the angular frequency and $c$ the speed of light. One possible set for the phase shifts of each diffraction order are
\begin{eqnarray}
\phi_0&=&0,\label{3p/strmatr/phase0_2}\\
\phi_1&=&-{1}/{2}\arccos{\left({(\eta_1^2-2\eta_0^2)}/{2\rho_0\eta_0}\right)},\label{3p/strmatr/phase1_2}\\
\phi_2&=&\arccos{\left({-\eta_1^2}/{(2\eta_2\eta_0)}\right)}\label{3p/strmatr/eq:12}.\label{3p/strmatr/phase2_2}
\end{eqnarray}
Since the phase shifts depend on the diffraction efficiencies one can derive a minimal/maximal diffraction efficiency for the zeroth and second diffraction order $\eta_{0/2}=(1\pm\rho_0)/2$. As a consequence the symmetric 3-port diffraction grating can be designed and manufactured with diffraction efficiencies only within these boundaries. However, independently from the grating's specific diffraction efficiencies the implementation of an additional power-recycling mirror with the amplitude coefficients $\rho_1$ and $\tau_1$ between the laser source and the grating forms two coupled resonators [see Fig.~\ref{Fig:3PGtoPR}~(c)]. In the following we will denote the first cavity, formed by the PR mirror and the grating, as the power-recycling cavity (PR cavity), with its parameters $\Phi_1$, $L_1$ and $d_1$, and the second cavity, formed by the grating and the end mirror, as the arm cavity, with its parameters $\Phi_2$, $L_2$ and $d_2$ given above. For such a system the amplitude reflection coefficients for the back-reflected field $c_{1\text{\tiny{PR}}}$, the forward-reflected field $c_{3\text{\tiny{PR}}}$, the intra-cavity field of the arm cavity $c_{2\text{\tiny{PR}}}$ and the transmitted field $t_{\text{\tiny{PR}}}$ are
\begin{eqnarray}
c_{1\text{\tiny{PR}}}&=&{\left[\rho_1-c_{1}\exp{(\mathrm{i}2\Phi_{1})}\right]}d_{1},\label{PR/c1}\\
c_{2\text{\tiny{PR}}}&=&{\mathrm{i}\tau_{1}\exp{\left[\mathrm{i}(\Phi_1+\Phi_{2})\right]}}c_{2}d_{1},\label{PR/c2}\\
c_{3\text{\tiny{PR}}}&=&{\mathrm{i}\tau_1\exp{(\mathrm{i}\Phi_{1}})}c_{3}d_{1},\label{PR/c3}\\
t_{\text{\tiny{PR}}}&=&{-\tau_1\tau_{2}\exp{\left[\mathrm{i}(\Phi_1+\Phi_{2})\right]}}c_{2}d_{1},\label{PR/t}
\end{eqnarray}
where $c_n$ are the amplitude reflection coefficients of the two-component cavity given above. The resonance factor of the PR cavity is given by $d_1=[1-\rho_{1}c_1\exp{(\mathrm{i}2\Phi_1})]^{-1}$, and the length $L_1$ of the PR cavity is expressed by the detuning parameter $\Phi_1=\omega{L_1/c}$.

In current gravitational wave detectors the arm cavities are standing-wave cavities with an amplitude reflectivity of the end mirror of $\rho^2_2\approx1$. Such a cavity always reflects the light back into the PR resonator, independently from the detuning of the cavity. This is in contrast to a 3-port-grating coupled cavity with the additional port C3. Hence, the light fields, that are back-reflected from the grating cavity and thus recoupled to the PR cavity, firstly depend on the grating-specific diffraction efficiencies and, secondly, on the tuning of the grating cavity. Note that the light amplitude reflected from the grating cavity $c_1$ (as given in Eq. (\ref{dreiport/c1})), can be seen as the compound component reflectivity $\rho_{\mathrm{c}}(\Phi_2)$ of the grating and the end mirror \cite{burmeister2010}. In consequence the overall power build-up inside both cavities finally depends on the detuning of the PR cavity $\Phi_{1}$ and the detuning of the arm cavity $\Phi_{2}$. The additional third port C3$\text{\tiny{PR}}$ is a loss channel showing a characteristic behavior which we present and discuss in the following experiment.

\section{Geometrical considerations}
\label{sub:geometrical_considerations}

Diffraction at a grating into an order $>0$ changes the geometrical parameters of a gaussian beam if the angle of incidence and the diffraction angle are different. This leads to restrictions to the geometrical set-up of a power-recycled 3-port-grating cavity for a stable resonator.
 
The arm cavity is a half-symmetric resonator consisting of a spherical end mirror (m) and a flat grating (g).
The two components form a stable resonator if the stability criterion is fulfilled:
\begin{equation}
	0~{\leq}~g_{\mathrm{g}}g_{\mathrm{m}}~{\leq}~1
\end{equation}
\begin{equation}
	\mathrm{with}\hspace{0.2cm} g_{\mathrm{g}}=1-\frac{L_2}{R_{\mathrm{c,g}}}=1,\hspace{0.2cm} g_{\mathrm{m}}=1-\frac{L_2}{R_{\mathrm{c,m}}},
	\end{equation}
where $R_{c,\mathrm{m}}$ denotes the radius of curvature of the end mirror and $L_2$ the length of the arm cavity as the second cavity of the system.
In a half-symmetric resonator the beam waist of the round eigenmode of the cavity is positioned at the flat component with a waist size
\begin{equation}
	w_0^2=\frac{L_2\lambda}{\pi}\sqrt{\frac{g_{\mathrm{m}}}{1-g_{\mathrm{m}}}}.
\end{equation} 

If a monochromatic TEM00 gaussian beam with round beam profile is incident on a grating in second-order Littrow configuration, the beam that is diffracted into the first diffraction order at an emergent angle of $0^\circ$ will be elliptical. The angle of incidence ($\theta_{\mathrm{in}}$) does not match the emergent angle ($\theta_{\mathrm{out}}$) which results in a distortion of the gaussian beam profile in one dimension. The relation between the waist size of input and output beam in the affected dimension is given by:
\begin{equation}
	w_{\mathrm{out}}=\frac{\cos(\theta_{\mathrm{out}})}{\cos(\theta_{\mathrm{in}})}w_{\mathrm{in}}.
\end{equation}  
Therefore the ratio between the x- and y-direction waist sizes is completely determined by the second-order Littrow angle (here $47.2^\circ$).
In order to match the mode of the incoming beam to the round eigenmode of the arm cavity, the incident beam needs to have a degree of ellipticity that exactly compensates the beam deformation that is applied to the beam when diffracted into the first order. As a consequence the radii of curvature of the PR mirror need to be different in x- and y- direction, because the eigenmode of the power-recycling cavity is elliptical. For a given eigenmode of the arm cavity and length of the recycling cavity $L_1$, the desired radii of curvature of the recycling mirror are given by: 
 \begin{eqnarray}
R_{\mathrm{c,y}} &=& L_1 + \frac{\pi^2w_{0,y}^4}{\lambda^2L_1},\\ 
R_{\mathrm{c,x}} &=& L_1 + \frac{\pi^2w_{0,x}^4}{\lambda^2L_1} = L_1 + \frac{\pi^2w_{0,y}^4}{\lambda^2L_1(\cos{\theta_{\mathrm{in}})^4}} .
 \end{eqnarray}

\begin{figure}[htbp]
	\centering
		\includegraphics[height=2.9cm]{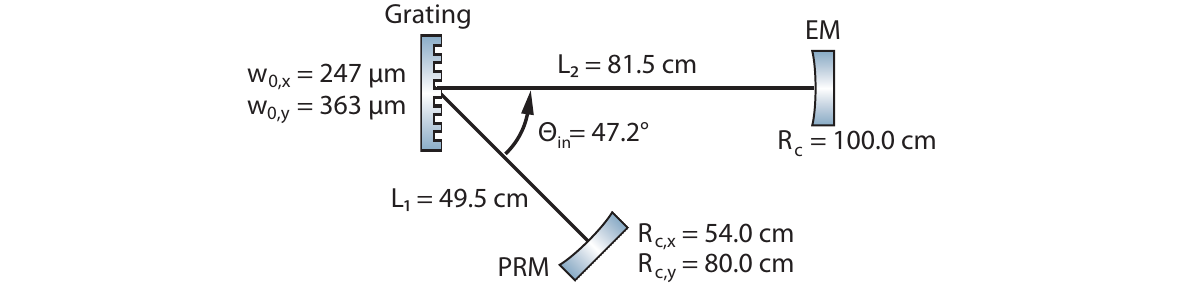}
	\caption{Geometrical configuration of the experiment. The second-order Littrow angle $\theta_{\mathrm{in}}$ determines the ratio of the waist size at the grating in horizontal and vertical direction $w_{\mathrm{0,x}}$ and $w_{\mathrm{0,y}}$. The length of the arm $L_2$ and the radius of curvature of the end mirror $R_{\mathrm{c}}$ defines the absolute waist size at the grating. Together with the length of the power-recycling cavity $L_1$ these parameters determine the radii of curvature of the PR mirror $R_{\mathrm{c,x}}$ and $R_{\mathrm{c,y}}$.}
	\label{fig:Bilder_topologie}
\end{figure}

The geometrical configuration of our experiment is shown in Fig.~\ref{fig:Bilder_topologie}. 

\section{Experimental set-up}

\begin{figure}[htbp]
	\centering
		\includegraphics[height=4.23cm]{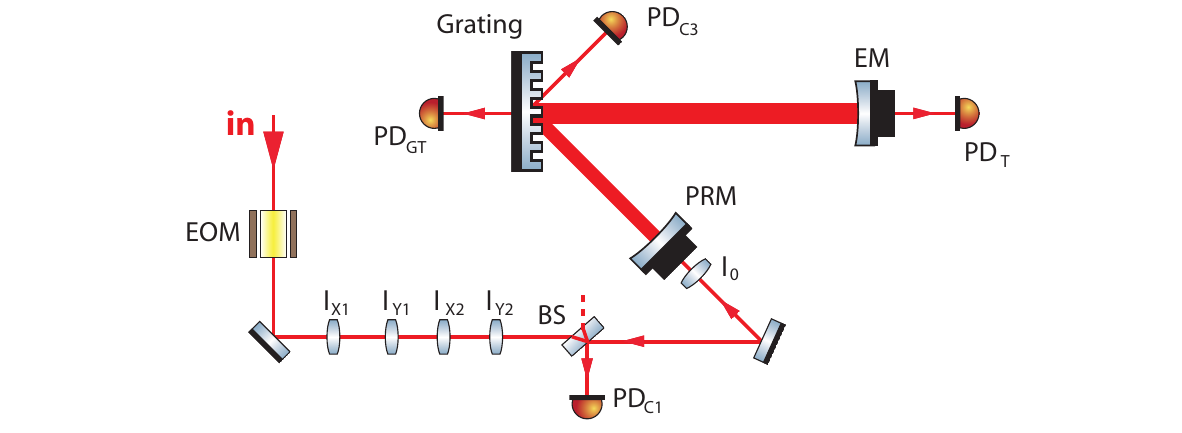}
	\caption{Experimental set-up of a 3-port-grating cavity with power-recycling. Two cylindrical lenses for each dimension l$_n$ are needed to provide mode-matching to the eigenmode of the cavity. The spherical lens l$_0$ compensates the distortion due to the PRM substrate. Photodiodes at the ports (PD$_{\mathrm{C}n}$) and in transmission of the grating PD$_\mathrm{GT}$ allow monitoring. The beamsplitter BS in the entrance allows access to the back-reflected field C1$_\mathrm{PR}$.}
	\label{fig:PR_setup}
\end{figure}

The experimental set-up is shown in Fig.~\ref{fig:PR_setup}. Light from a Nd:YAG laser was spatially filtered by a ring-mode-cleaner \cite{PMC}. An electro-optical modulator (EOM) generated a 15\,MHz phase modulation for a Pound-Drever-Hall (PDH) lock \cite{drever} of the cavity lengths. We designed and fabricated \cite{clausnitzer05} a binary dielectric reflection 3-port grating with close to minimal second-order diffraction efficiency. We chose a grating period of 1450\,nm, which provided a second-order Littrow angle of $\theta_{\mathrm{in}}=47.2^\circ$ and a first oder diffraction angle of $\theta_{\mathrm{out}}=0^\circ$ at the laser wavelength of 1064\,nm. For each dimension two cylindrical lenses were employed to generate the elliptical beam profile that is needed in order to mode-match the incoming beam to the eigenmode of the recycling cavity. A dieletric multilayer system was applied to the PR mirror providing a reflectivity of $\rho_1^2=0.96$. The high-reflectivity end mirror had a transmissivity of $\tau_2^2=7\,$ppm. The PR mirror as well as the end mirror were mounted onto a piezo-electrical transducer (PZT) to linearly sweep the cavity lengths. 

Initially the grating was characterized via a finesse measurement using the set-up discussed in \cite{Bunki06}. The diffraction efficiencies of the grating for s-polarized light were found to be $\eta_0^2=0.927\pm0.045$, $\eta_1^2=0.0591\pm0.003$ and $\eta_2^2=0.0001\pm50\,$ppm and the reflectivity of the grating for normal incidence was $\rho_0^2=0.879\pm0.003$. The total optical loss of the grating due to scattering, absorption and residual transmission was determined to $A=0.0027\pm0.0061$. The grating transmission was independently measured and was found to be $\tau_0^2=123\,$ppm. We used the light transmitted through the grating to monitor the power inside the arm cavity.  

When the PR mirror was misaligned with respect to the actual optical axis of the recycling cavity (e.g.~tilted), the mirror acted as an attenuator in front of the arm cavity. In this configuration the arm cavity properties could be determined without being influenced by the recycling cavity. The finesse of the arm cavity was found to be $\mathcal{F}_{\mathrm{arm}}=49.3$ deduced from the measured line-width of $\mathrm{FWHM}_{\mathrm{arm}}=3.73\,\mathrm{MHz}$. In a second step we re-aligned the PR mirror and locked the arm cavity length on resonance using the standard PDH method. We then scanned the length of the power-recycling cavity and also determined the finesse of the recycling cavity. Note that in this configuration the stabilized arm cavity acted as compound cavity end mirror with reflectivity $\rho_{\mathrm{c}}(\Phi_2=0)$ \cite{burmeister2010}. The measured linewidths of the recycling cavity of $\mathrm{FWHM}_\mathrm{PR}=5.79\,$MHz yielded a finesse of $\mathcal{F}_\mathrm{PR}=52.3$. 

\section{Power-recycling and cavity detunings} 
\label{sub:comparison_with_simulations}

In this section we present measurements of the light power inside the arm cavity and at the two output ports of the coupled cavity system versus the detunings of the arm cavity and the PR cavity, respectively. We compare our results with a model that is based on Eqs.~(1--11) and measured diffraction efficiencies, mirror reflectivities and losses of the optical components. Note that our theoretical description of the 3-port-grating assumes zero optical loss \cite{bunkowski1}. In order to account for grating loss our model used a transmission of the arm cavity end mirror that was slightly higher than the actually measured one. Nevertheless, quantitative differences between our model and our experiment are still expected due to the imperfect matching of input and cavity modes, respectively.  

Generally, the state of the coupled optical system depends on the detuning of both cavities $\Phi_1$ and $\Phi_2$. We therefore present the results of our model as three-dimensional plots showing light powers versus detunings \cite{burmeister2010}. However, in our measurement the detunings were not stabilized to certain points of the phase space but swept by applying sinusoidal voltages to piezo-electric transducers behind the cavity mirrors. One of the cavity detunings was varied slowly with a frequency below 1\,Hz, whereas the other was varied fast with a frequency of about 1\,kHz. From this procedure we derived two-dimensional plots showing light power versus the one detuning that was varied slowly. However, for any value of the slowly varied detuning, the voltage from the respective photo diode was sampled fast enough to pick its extreme values that occurred due to the fast varied detuning. The measured two-dimensional plots therefore correspond to a projection of our three-dimensional plot onto one of the detuning axes thereby collecting all maxima and minima along the other axis. The theoretical pictures for comparison with the experimental data corresponds to `side-views' onto the plain in the three-dimensional plot along one or the other detuning axis.

\begin{figure}[hbtp]
	\centering
		\includegraphics[height=7.5cm]{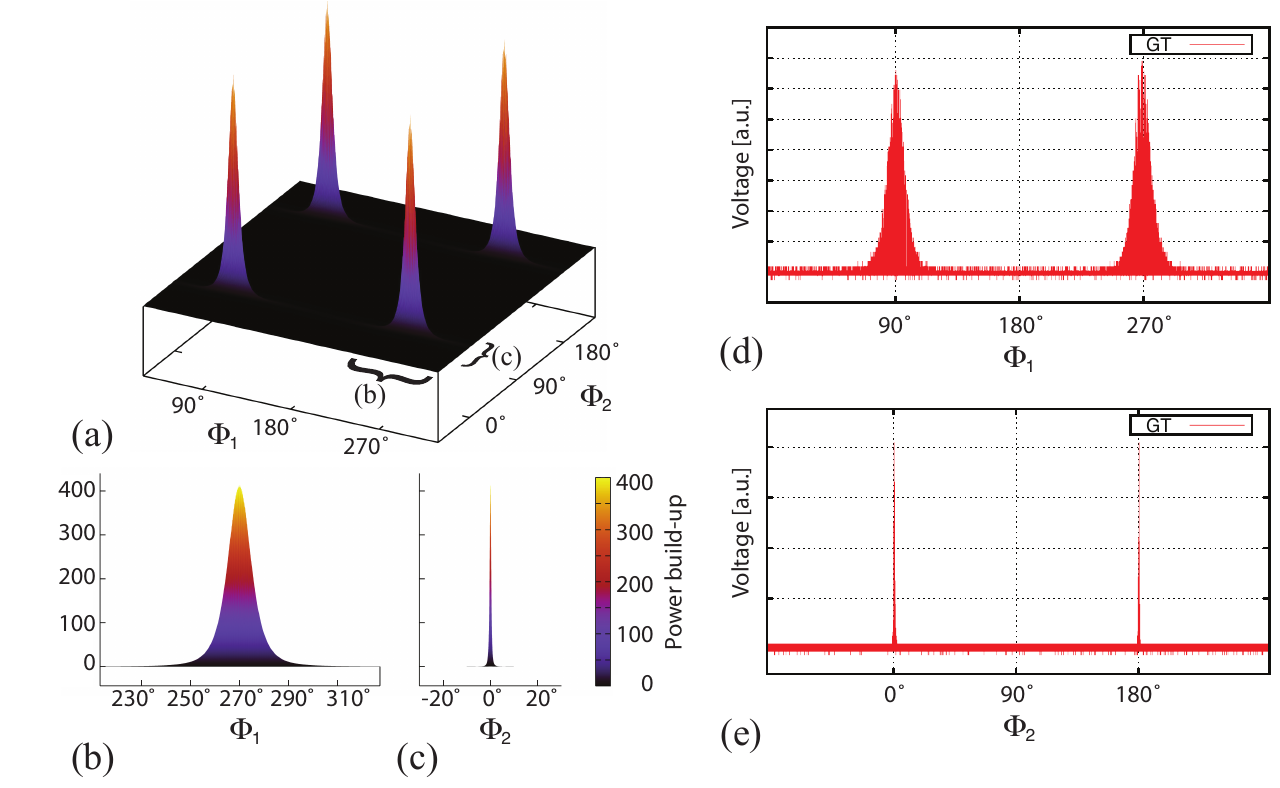}
	\caption{Intra-cavity power build-up in the arm cavity as a function of the detunings $\Phi_1$ and $\Phi_2$ of the PR cavity and the arm cavity, respectively. Plot (a) shows the 2-dimensional simulation. Plots (b) and (c) are projections onto one of the axes, respectively. They well reproduce the experimental data as given in plots (d) and (e). The latter two show the intra-cavity powers transmitted through the grating as detected by photo diode PD$_\mathrm{GT}$. For both plots one of the cavity lengths was varied slowly whereas the other was varied fast. 	
	}
	\label{fig:GT}
\end{figure}

Figure \ref{fig:GT} presents the simulated values of the intra-cavity power versus detunings and the light power as detected by photo diode PD$_\mathrm{GT}$ when varying the detunings in the way described before. The resonance pattern of Fig.~\ref{fig:GT}\,(a) is periodic in $\Phi_1\,\mathrm{mod}\,\pi$ and in $\Phi_2\,\mathrm{mod}\,\pi$. Power enhancement is present around a maximum at $\Phi_1=90^\circ$ and $\Phi_2=0^\circ$. These phases are a consequence of the grating phase relations as introduced earlier, and account for 3-port gratings with a (close to) minimal value for the second order diffraction efficiency $\eta^2_2$ \cite{burmeister2010}. Simulated and experimental data in Fig.~\ref{fig:GT}\,(b,c) and (d,e), respectively, show a very good qualitative agreement.
The existence of isolated, equidistant resonances in Figure \ref{fig:GT} is a special property of the optical cavity system as investigated here, and is in contrast to a conventional three-mirror cavity (confer Refs.~\cite{burmeister2010,Thuering}). It is in fact a direct consequence of the additional port C3 (Fig.~\ref{fig:PR_setup}). In order to gain a high power build-up in the arm cavity, output at this port has to be avoided. This happens only if the detunings of the subsystem cavities have particular values (modulo $\pi$) providing destructive interference between the field from the power-recycling mirror and the field leaking out the arm cavity. 

A stable operation of our cavity system at arbitrary points in phase space was not realized, however, it was realized at the point of maximum power build-up in the arm cavity. The error signal for the PDH length stabilization of the power-recycling cavity was generated from the output of photo-diode PD$_\mathrm{C1}$. The light power on photo-diode PD$_\mathrm{GT}$ and the known transmissivity of the grating then allowed to determine the light power inside the arm cavity. This value turned out to be at least 16 times higher than the independently measured power build-up inside the arm cavity without power-recycling, both normalized to the same input light power. This value was slightly smaller than our simulated value of $\approx24$ due to imperfect matching of the light's input mode and the cavity modes as well as due to the finite bandwidth of the cavity locking loops. 

\begin{figure}[htbp]
	\centering
		\includegraphics[height=7.5cm]{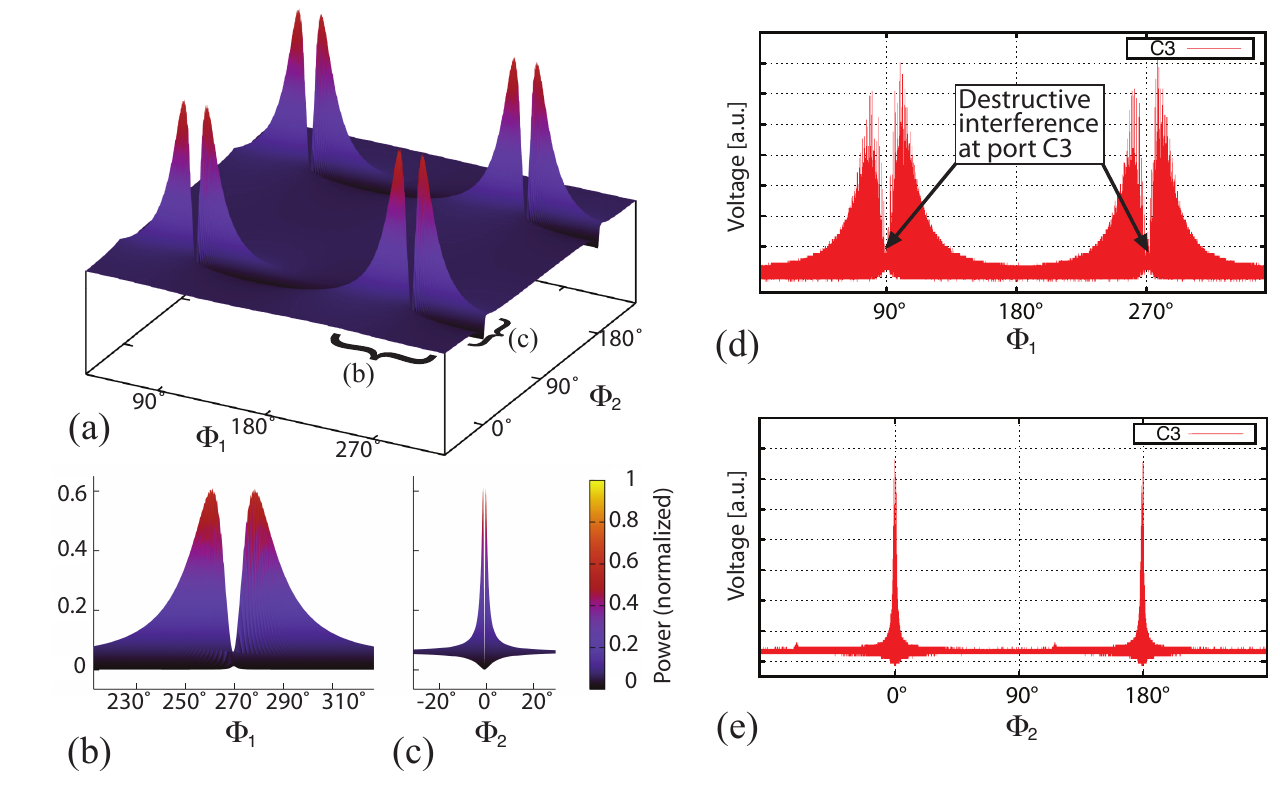}
	\caption{Output power at the additional grating port C3$_\mathrm{PR}$ as a function of the detunings $\Phi_1$ and $\Phi_2$ of the power-recycling cavity and the arm cavity, respectively. Plot (a) shows the 2-dimensional simulation. Plots (b) and (c) are projections onto one of the axes, respectively. They well reproduce the experimental data as given in plots (d) and (e). The latter two show the output powers as detected by photo diode PD$_\mathrm{C3}$. For both plots one of the cavity lengths was varied slowly whereas the other was varied fast. The different peak heights in Fig.~(d) were due to non-linearities in the PZTs used to vary the cavity lengths.
	}
	\label{fig:C3}
\end{figure}

Fig.~\ref{fig:C3} presents the characterization of the additional port C3 of our cavity system. It shows simulated and measured light power leaving the coupled system at port C3. Again simulated and measured data show a high degree of agreement. In Fig.~\ref{fig:C3}\,(a) it can be seen that for the detunings that provide the highest power build-up in the arm cavity a minimum light power is coupled out via port C3. Due to our measurement procedure this minimum is visible only in the projection onto detuning $\Phi_1$. However, stabilizing both cavities to the phase space point of maximum power inside the arm cavity, as described above, clearly provided a minimum light power at port C3. The doublet structure of the resonance in Fig.~\ref{fig:C3} is due to the power-recycling effect. When both cavities are far detuned from resonance most of the light is back-reflected towards the laser source. If the arm cavity is on resonance considerable power build-ups occur for a relatively wide range of detunings of the power-recycling cavity. But only for the optimum detuning of the power-recycling cavity ($\Phi_1=90^\circ$) destructive interference occurs in port C3 giving rise to the sharp dip as shown.

\section{Conclusion}
We have demonstrated the diffractive coupling of two optical resonators. In full agreement with our theoretical work in \cite{burmeister2010} we have experimentally shown that a dielectric 3-port grating can replace the transmissive coupling mirror located between a power-recycling cavity and a Fabry-Perot arm cavity as used in gravitational wave detectors. In combination with an all-reflective balanced beam splitter as demonstrated recently \cite{gratingbs}, a power-recycled Michelson interferometer is feasible without having to transmit high powers through internal optical components, as shown in Fig.~\ref{Fig:IFO_heat}. Although a 3-port grating as a coupling component of two resonators opens an additional port, which is the third grating port C3, a significant power-recycling factor was achieved. Almost no power was lost into the additional port due to destructive interference if the detunings of the two resonators were chosen properly. Since a 3-port grating introduces elliptical beam profiles the mode-matching of the coupled cavities was challenging; however, it could be efficiently realized utilizing a PR mirror with two different radii of curvature in horizontal and vertical dimension. A 3-port grating combines three, instead of the two light fields at each port. We have carried out a simulation based on measured parameters of the optical components and found good qualitative agreement between simulated and measured data. We conclude that all-reflective optics are a promising technology for future laser interferometric gravitational wave detectors, where coupled optical cavities are routinely used to enhance the stored laser power in the arms to improve the shot noise limited sensitivity. 

\section*{Acknowledgements} This work has been supported by Deutsche Forschungsgemeinschaft within the Sonderforschungsbereich TR7 and the Excellence Cluster QUEST, the IMPRS on Gravitational Wave Astronomy and the ET design study supported by the European Commission under FP7 (Grant Agreement 211743).

\end{document}